\documentclass[aps,prl,twocolumn,superscriptaddress]{revtex4}
\usepackage{amsfonts}
\usepackage{amsmath}
\usepackage{amssymb}
\usepackage{graphicx}

\setcitestyle{super}

\begin{document}
\title{A simple diatomic potential that prevents crystallization in supercooled liquids simulations}


\author{A. P. Kerasidou}
\affiliation{ Laboratoire de Photonique d'Angers EA 4464, University of Angers, Physics Department,  2 Bd Lavoisier, 49045 Angers, France.}
\affiliation{MOLTECH Anjou, University of Angers,  2 Bd Lavoisier, 49045 Angers, France.}

\author{Y. Mauboussin}
\affiliation{MOLTECH Anjou, University of Angers,  2 Bd Lavoisier, 49045 Angers, France.}

\author{V. Teboul*}
\affiliation{ Laboratoire de Photonique d'Angers EA 4464, University of Angers, Physics Department,  2 Bd Lavoisier, 49045 Angers, France.}
\email{victor.teboul@univ-angers.fr}


\begin{abstract}

We study a simple and versatile diatomic potential function coined to prevent crystallization in supercooled liquids. 
We show that the corresponding liquid doesn't crystallize even with very long simulation runs at the lowest temperature that we can access with ergodic simulations. The medium displays the usual features of supercooled materials and a non-Arrhenius dependence of the diffusion coefficient and $\alpha$ relaxation time with temperature.
We also observe the breakdown of the Stokes-Einstein relation at low temperatures.


\end{abstract}

\maketitle
\includegraphics[scale=0.15]{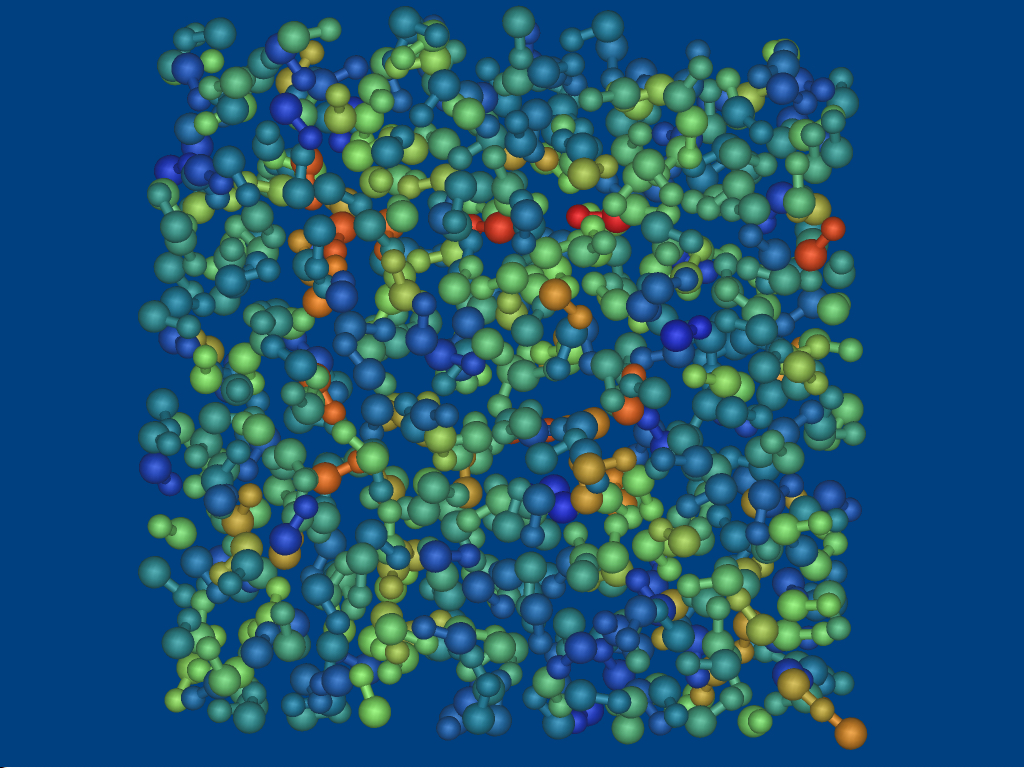}


\section{Introduction}
Supercooled liquids undergo an exponential (Arrhenius) or even larger increase of their viscosity when the temperature decreases.
This large modification of the transport properties appear while the structure changes only slightly with temperature.  
While several theories\cite{gt2,gt3} have been proposed to solve that long standing glass-transition problem, it is still open\cite{anderson,gt1,gt2,gt3,gt4}.
Interestingly while the reasons for the strange behavior of supercooled liquids are still not understood, molecular dynamics simulations\cite{md1} reproduce the unexplained phenomena\cite{gt1,gt2}.
Consequently molecular dynamics simulation is an invaluable tool\cite{md1,visco,prl,w1,w2,w3} to study the glass-transition problem, and more generally\cite{md1,md3,md4,md5,md6,cage,md7} for the study of condensed matter physics.
Due to the universality of the glass-transition  phenomenology\cite{gt1,gt2,gt3,gt4} one is tempted to chose the simplest existing potential function in order to simplify as much as possible the complexity of the problem.
Unfortunately when the potential is too simple, the liquid crystallizes rapidly.
Thus one is conducted to search for the simplest potential that prevents crystallization for supercooled liquids simulations while displaying as strongly as possible the dynamical behavior of supercooled liquids. Various potentials\cite{potG,potDz1,potDz2,potDz3,potKL1,potKL2,potKA1,potKA2} have been proposed in that purpose. However some eventually undergo partial crystallization for long runs, while others are not so simple. 
The most popular potential to date that hinders the crystallisation is the Kob-Andersen potential\cite{potKA1,potKA2}. The corresponding liquid is a mixture of two different Lennard-Jones atoms (A and B)  with a proportion of $20\%$ of atoms A and $80\%$ of atoms B. 
If that potential is one of the simplest it also creates  an unnecessary increase of the complexity of the problem due to the mixture of atoms.
However liquids constituted with only unmixed atoms A or B do crystallize very fast.
A simple idea to overcome that problem is to bound the two atoms A and B creating a diatomic molecule which liquid may be expected not to crystallize. Unfortunately after long runs that liquid also crystallizes partially.

In this work we study a simple and relatively versatile potential function, based nonetheless on that idea of two bounded Lennard-Jones atoms which parameters are chosen to prevent crystallization. Due to its simple Lennard-Jones structure we expect that potential to be a good candidate to model the universal physics of molecular liquids.
We show that the liquid constituted by these molecules doesn't crystallize at low temperatures and that it follows the typical behavior of non-Arrhenius supercooled liquids.

\section{Calculation}

We model the molecules as constituted of two atoms ($i=1, 2$) that do interact with the following Lennard-Jones potentials: 
$V_{ij}=4\epsilon_{ij}((\sigma_{ij}/r)^{12} -(\sigma_{ij}/r)^{6})$ with the parameters: $\epsilon_{11}= \epsilon_{12}=0.5 KJ/mol$, $\epsilon_{22}= 0.4 KJ/mol$, $\sigma_{11}= \sigma_{12}=3.45$\AA\ and $\sigma_{22}=3.28$\AA$ $.
Note that as in the Kob-Andersen model\cite{potKA1,potKA2} we do not use the usual additive mixing rules\cite{md1} for Lennard-Jones potentials. 
We make that choice of non-additive mixing rules as it has the property to hinder the crystallization and the formation of plastic crystal phases\cite{pc1,pc2,pc3,M1,M2,M3,M4}.
We use the mass of Argon for each atom of the linear host molecule that we rigidly bonded fixing the interatomic distance to $d=1.73 $\AA$ $.  
The reduced shape\cite{M1,M2} of our dumbbell molecule is $L^{*}=d/\sigma=0.5$ a value somehow larger than the limit $L^{*}=0.4$ below which plastic phases are usually created\cite{M1,M2}.
With these parameters the liquid does not crystallize even during long run simulations.
The simulations are first equilibrated during $20$ to $100 ns$ depending on the temperature, then we perform the production run.
 However a few long runs  of $400 ns$ each, have also been realized at low temperature ($T=40 K$) with a smaller simulation box containing $500$ molecules only and we didn't find any sign of crystallization.
We use the Gear algorithm with the quaternion method\cite{md1} to solve the equations of motions with a time step $\Delta t=10^{-15} s$. The temperature is controlled using a Berendsen thermostat\cite{berendsen}. The density is set constant at $\rho=2.24 g/cm^{3}$. 
We use in our calculations a cubic simulation box that contains $N=2688$ molecules and has a length $L= 53 $\AA.
The model has the interesting property to be versatile allowing modifications without  crystallizing.
For example one can easily change the mass of the atoms leading to different densities or change the interatomic distance $d$.

\section{Results and discussion}

To verify that there is no crystallization in our liquid, we plot in Figure 1 the radial distribution function (RDF) for various temperatures ranging from above the melting temperature to a deep supercooled liquid. 

\includegraphics[scale=0.33]{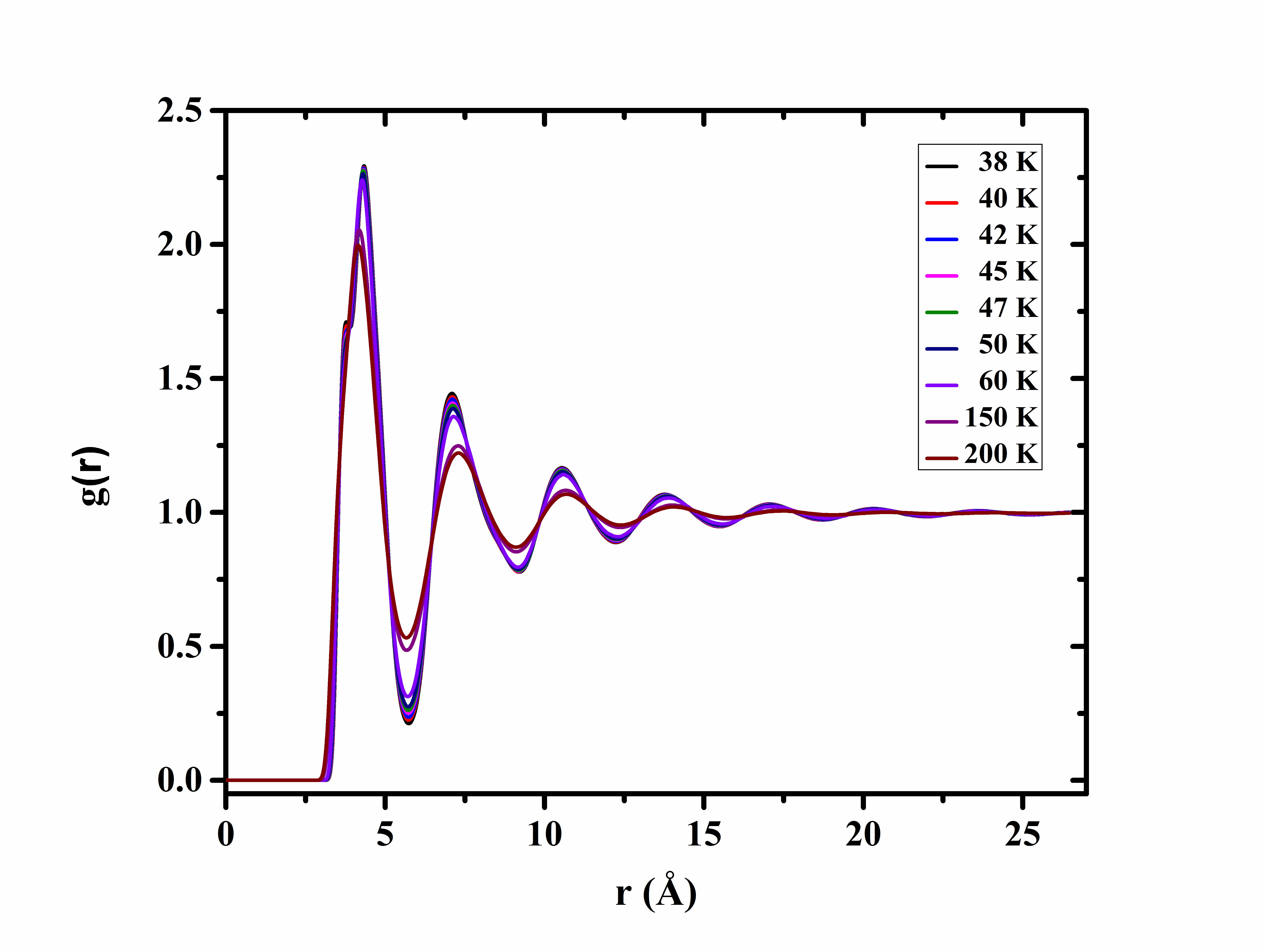}

{\em \footnotesize  FIG.1. (color online) Radial distribution function $g(r)$ between the molecules center of masses for various temperatures. The structure doesn't change much even at low temperature. We do not see any peak signalling a partial crystallization at low temperatures. Below 150K the liquid is supercooled.\\}

 The radial distribution function $g(r)$ represents the distribution probability to find a molecule a distance $r$ apart from another molecule.
We see on the Figure that the RDF doesn't change much when the temperature decreases from above the melting temperature down to the lowest accessible temperature with our simulations ($T=38 K$). As the temperature decreases, the peaks increase in size, but we do not see any modification of the maxima and minima locations that would have been the signature of a modification of the structure. We do not see ever any sharp peak signalling a partial crystallization of the liquid. Moreover the structure appears to be very simple, quite like the structure of a simple monatomic liquid. The main difference is a small shoulder that we observe in the first peak of the RDF. This shoulder appears due to the difference between the Lennard-Jones potentials of the two atoms constituting the molecule.

\includegraphics[scale=0.33]{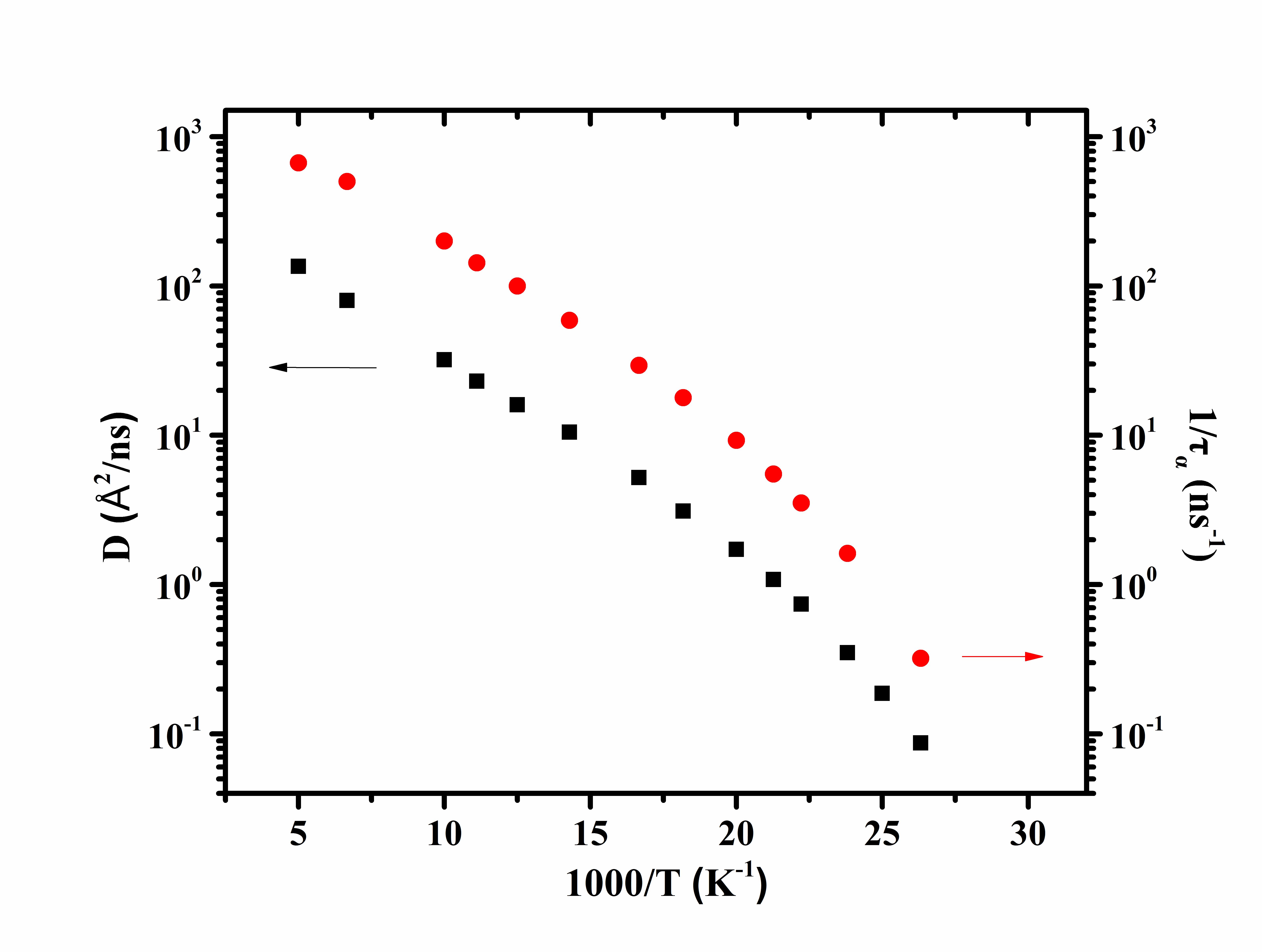}

{\em \footnotesize   Fig.2. (color online) Diffusion coefficient (black squares) and inverse of the $\alpha$ relaxation time (red circles) evolution with temperature. The Figure shows an evolution faster than a pure exponential. That evolution called super-Arrhenius is typical of molecular supercooled liquids.\\}

We will now study the dynamic properties of the liquid.
The diffusion coefficient as well as the $\alpha$ relaxation time in Figure 2 display a super-Arrhenius evolution with temperature.
If the diffusive motions are thermally activated processes we may write: $D=D_{0}.e^{-E_{a}/k_{B}T}$, where $E_{a}$ is the activation energy to overpass for the molecule to diffuse. 
A super-Arrhenius evolution implies that the activation energy $E_{a}$ is increasing when the temperature decreases and for that reason that behavior has been associated in the past with the emergence of cooperativity. 
This super-Arrhenius behavior also show that our liquid is a "fragile" liquid in Angell's classification\cite{angell1,angell2}.
This comportment is typical of molecular liquids, suggesting that our model is well suited for molecular liquids.
We also observe that the $\alpha$ relaxation time evolves more rapidly than the diffusion coefficient at low temperatures.
This result suggests a breakdown of the Stokes-Einstein relation\cite{stokes-einstein}.
Note that the diffusion coefficient $D$ decreases continuously with the temperature showing no sign of a sharp drop of $D$ induced by a crystallization process. 

\includegraphics[scale=0.33]{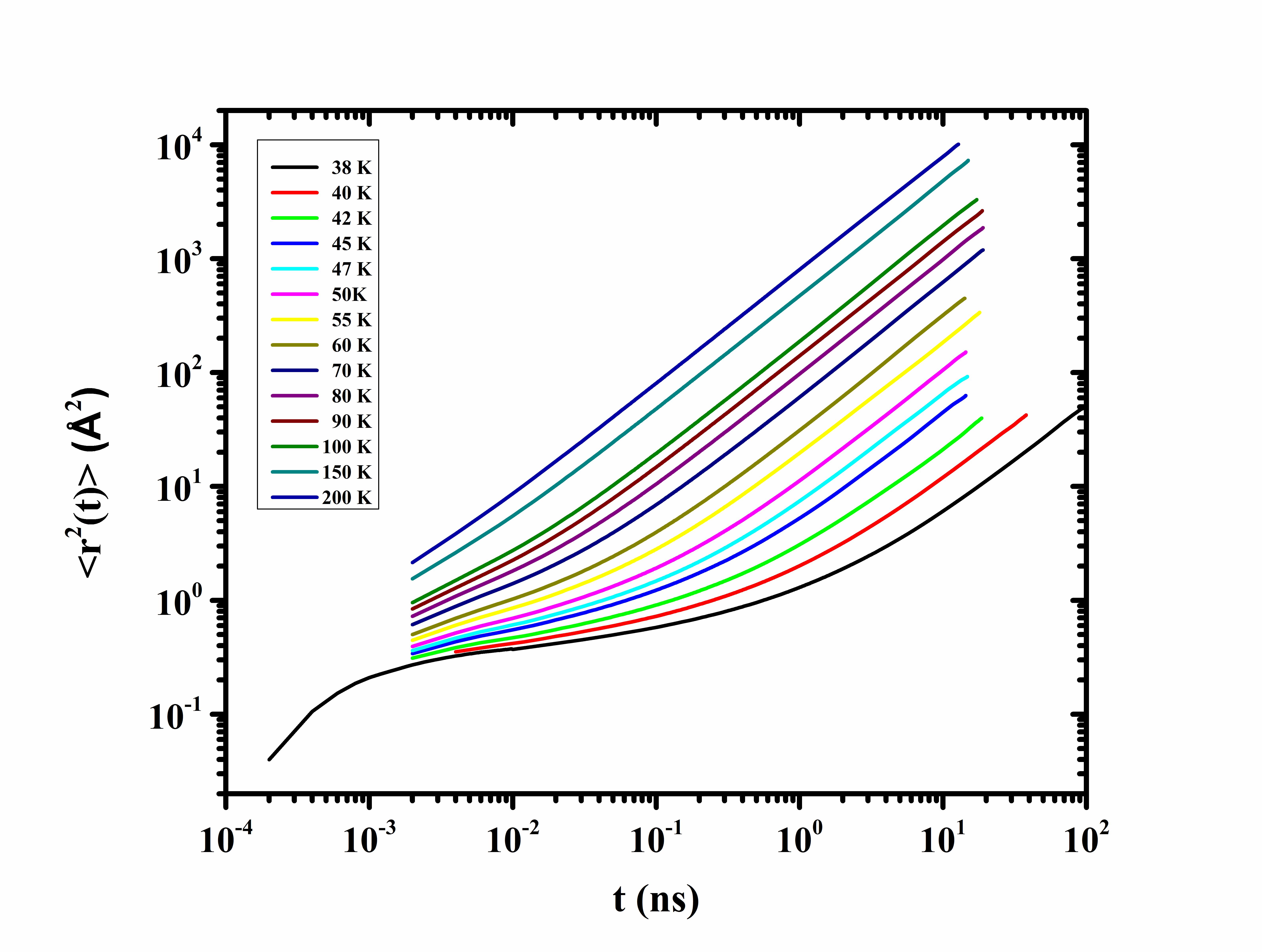}

{\em \footnotesize  FIG.3. (color online) Mean square displacement of the molecules center of masses plotted for various temperatures. The curves display a plateau typical of the supercooled state and that disappears above the melting temperature $T_{m}\approx150 K$.\\}

Figure 3 shows the mean square displacement of the molecules for various temperatures. 
We observe the three different time regimes characteristic of supercooled liquids.
For short time scales ($t<10^{-3} ns$) the molecules are in the ballistic regime, between $10^{-3} ns$ and $1 ns$ for the smaller temperature investigated (black curve, T=38 K) we are in the plateau regime and for time scales larger than $1 ns$ we observe the diffusive regime.
The plateau regime appears for temperatures below $T\approx150K$.
The appearance of a plateau is characteristic of supercooled liquids, and we deduce from that emergence that the melting temperature $T_{m}\approx150 K$ in our model liquid. The plateau in the mean square displacement is a signature of the cage effect\cite{cageM} and corresponds to the mean time lapse during which the molecule is trapped inside the cage constituted by its neighbors.
The width of the plateau increases when the temperature drops leading to the increase of the relaxation times.
The diffusion coefficient $D$ plotted in Figure 2 is calculated from these curves using the relation: 

$\displaystyle \lim_{t \to \infty} <r^{2}(t)>=6 D t$

\includegraphics[scale=0.33]{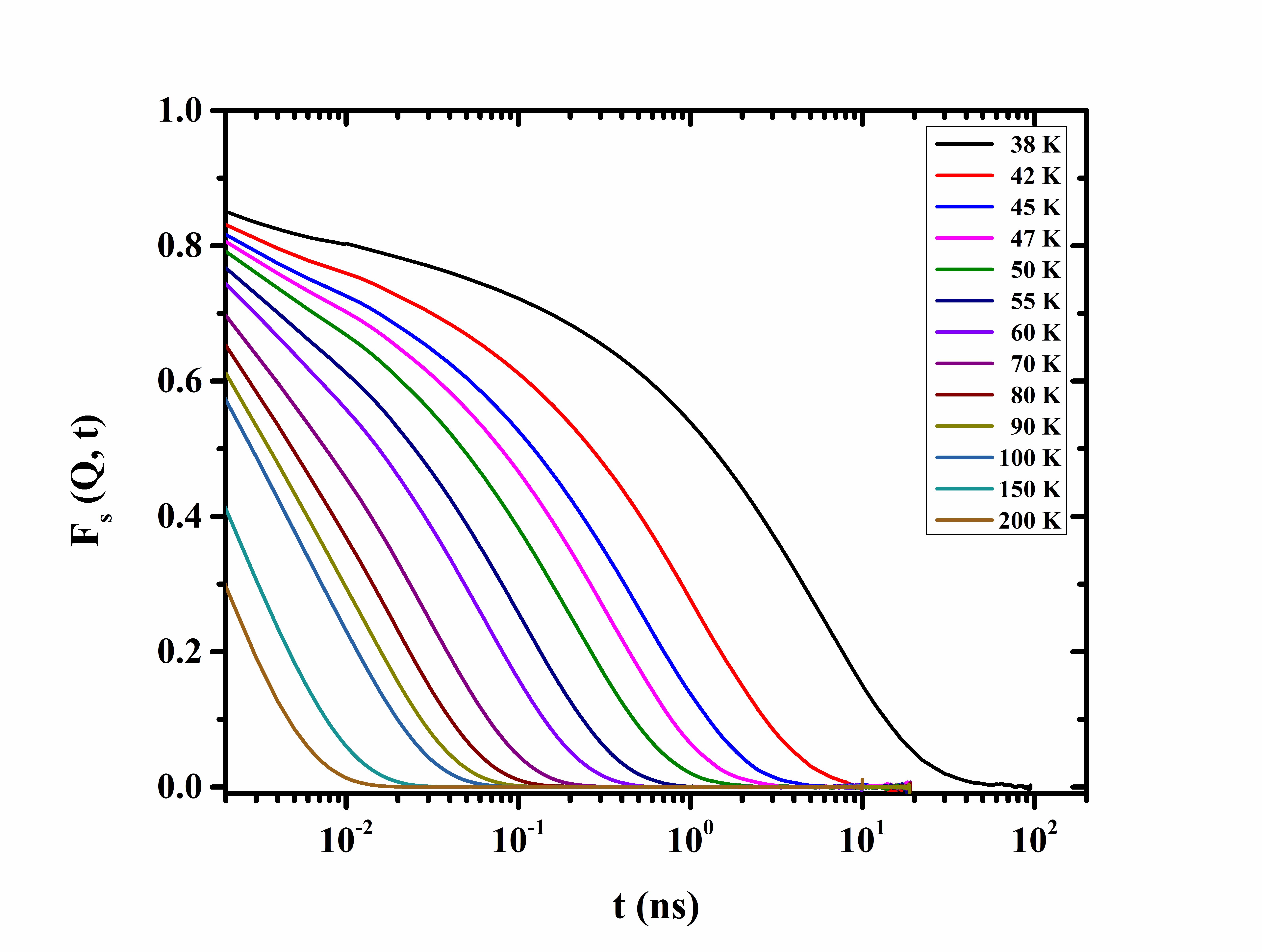}

{\em \footnotesize  FIG.4. (color online) Incoherent intermediate scattering function $F_{S}(Q,t)$ for the molecules center of masses at various temperatures. $F_{S}(Q,t)$ is calculated for a wave vector $Q=1.9$\AA$^{-1}$ that corresponds to the location of the first peak of the structure factor $S(Q)$. The functions are normalized so that  $F_{S}(Q,t=0)=1.$The rapid decrease of the very beginning of the curves displaying the ballistic motion of molecules inside their cages is not shown.\\}

Another important correlation function is the incoherent intermediate scattering function, defined as:  $F_{S}^{com}({\bf Q},t)=\displaystyle{ <{1\over N} \sum_{i=1}^{N}{e^{i{\bf Q}.({\bf r}_{i}(t+t_{0})-{\bf r}_{i}(t)) } >}}$. 
For an amorphous medium, any direction is equivalent and we may replace the wave vector $\bf Q$ by its modulus. 
The incoherent intermediate scattering functions $F_{S}(Q,t)$ are displayed in Figure 4 for various temperatures. As for the mean square displacements in the previous Figure, we observe  the appearance below $T_{m}$ of a plateau that increases when the temperature drops.
The function $F_{S}(Q,t)$ displays the typical evolution of the incoherent intermediate scattering function for supercooled liquids.
We also observe the stretching of the functions for the $\alpha$ relaxation (i.e. the functions do not decrease with a simple Debye relaxation law).
The stretching of the incoherent scattering function has been associated in the past to the appearance of cooperative motions (dynamical heterogeneities) in supercooled liquids, as  that stretching can easily be interpreted as arising from the superposition of different sort of relaxations with different $\alpha$ relaxation times.
The $\alpha$ relaxation time ${{\tau}_{\alpha}}$ plotted in Figure 2 is calculated from these curves using the relation: $F_{S}(Q,{{\tau}_{\alpha}})=e^{-1}$.

\includegraphics[scale=0.3]{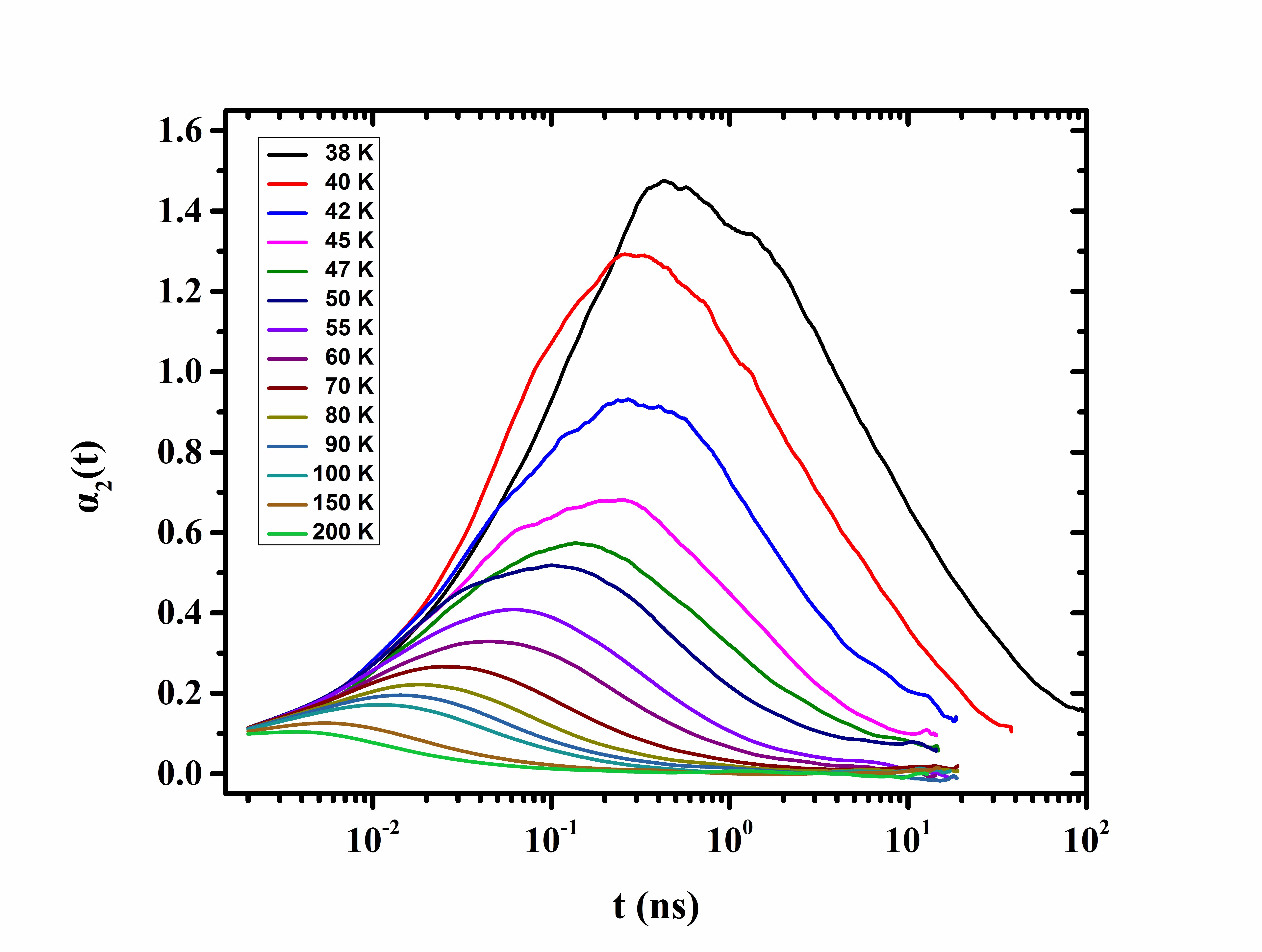}

{\em \footnotesize  FIG.5. (color online) Non-Gaussian parameter $\alpha_{2}(t)$ versus time for different temperatures. The maximum value of the non-Gaussian parameter as well as its characteristic time $t^{*}$ increase when the temperature drops. This behavior is induced in supercooled liquids by the appearance of spontaneous cooperative motions called 'dynamic heterogeneity'. We define $t^{*}$ from the relation: $\alpha_{2}(t^{*})=\alpha_{2}^{max}$.\\}

Figure 5 shows the evolution of the non-Gaussian  parameter $\alpha_{2}(t)$ (NGP) when the temperature drops. 
The non-Gaussian parameter defined as: $\alpha_{2}(t)=\displaystyle{{3\over 5}{<r^{4}(t)>\over<r^{2}(t)>^{2}} -1}$, measures the deviation of the van Hove correlation function from the Gaussian shape required by Brownian motions.
The main reason for this deviation at low temperature is that the Van Hove develops a tail corresponding to molecules moving cooperatively in motions larger than the average. 
As a result $\alpha_{2}(t)$ is, in most cases, a measure of cooperativity in supercooled liquids.
As the temperature drops the maximum value of the NGP increases as well as its characteristic time $t^{*}$ defined from the relation: $\alpha_{2}(t^{*})=\alpha_{2}^{max}$.
This result, typical of supercooled liquids, suggests the presence of cooperative motions in the liquid, that do increase when the temperature drops. 

\includegraphics[scale=0.33]{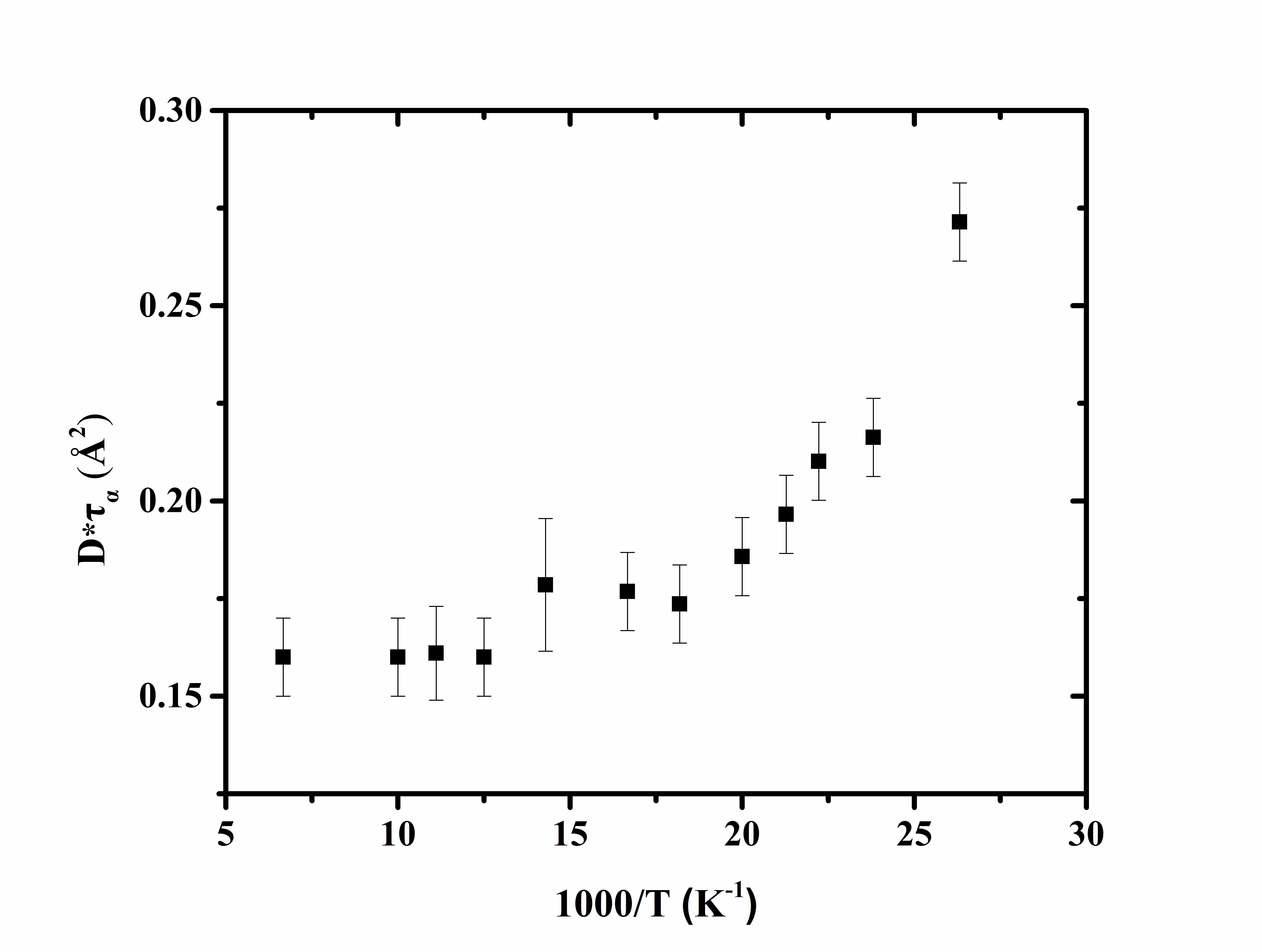}

{\em \footnotesize  FIG.6. (color online) Breakdown of the Stokes-Einstein relation at low temperatures. We see on the Figure that $D.\tau_{\alpha}$ is constant at high enough temperatures, showing that the Stokes-Einstein relation holds for that range of temperatures.  At lower temperatures however we observe a breakdown of the relation as  $D.\tau_{\alpha}$ increases rapidly. This behavior is usually associated with the appearance of cooperativity.\\}

Another test of the appearance of  cooperative motions in supercooled liquids is the breakdown of the Stokes-Einstein relation. 
The relation stands that the diffusion coefficient $D$ is inversely proportional to the viscosity $\eta$ of the medium.
 More precisely $D=k_{b}T/( 6\pi  \eta a)$ where $a$ is the characteristic size of the particles (here the molecules) that diffuse.
 As \cite{stokes-einstein} $\tau_{\alpha} \approx \eta/T$ the Stokes-Einstein relation leads to $D. \tau_{\alpha} = constant$.
However for supercooled liquids there is a breakdown of the relation due to the appearance of cooperative motions in the liquid.
This is the behavior that we do observe in Figure 6.
$D. \tau_{\alpha}$ stays approximately constant for $1000/T<20$ $ K^{-1}$ i.e. for $T>50$ $ K$, then it increases, showing that the diffusion coefficient $D$ is decreasing more slowly than  the $\alpha$ relaxation time $\tau_{\alpha}$ (i.e. the viscosity)  increases.

\section{Conclusion}

In summary we have implemented a very simple potential model that doesn't crystallize at low temperatures.
We have shown that this potential reproduces the structural and dynamical behaviors of typical supercooled glass-formers and in particular the appearance of cooperative motions that increase when the temperature drops. 
While  these behaviors are universal, some glass-formers do have more pronounced cooperative motions than others making them better candidates as a model, and we have shown that the simple intermolecular potential considered in our study is in that viewpoint a good candidate to model supercooled molecular liquids.


\begin{thebibliography}{99}


\bibitem{gt2} P.G. Wolynes, V.  Lubchenko, 
\newblock  {\em  Structural Glasses and Supercooled Liquids}, Wiley, Hoboken 2012

\bibitem{gt3} L. Berthier, G.  Biroli, J.P.  Bouchaud, L. Cipelletti, W. Van Saarlos, 
\newblock  {\em  Dynamical Heterogeneities in Glasses, Colloids and Granular Media}, Oxford Univ. Press, Oxford 2011

\bibitem{anderson} P.W. Anderson,  \emph{Science} \textbf{267},
1615 (1995)



\bibitem{gt1} K. Binder,  W. Kob,
\newblock  {\em  Glassy Materials and Disordered Solids}, World Scientific, Singapore 2011


\bibitem{gt4} P.G. Debenedetti, 
\newblock  {\em  Metastable Liquids}, Princeton Univ. Press, Princeton 1996

\bibitem{md1} M.P. Allen, D.J. Tildesley, 
\newblock  {\em  Computer simulation of liquids}, Oxford University Press, New York 1990


\bibitem{visco} R. Yamamoto, A. Onuki  \emph{Phys. Rev. Lett.} \textbf{81},
4915 (1998)

\bibitem{prl}  V. Teboul, M. Saiddine, J. M. Nunzi, 
\newblock  {\em Phys. Rev. Lett. } {\bf  103}, 265701 (2009)

\bibitem{w1}  V. Testard, L. Berthier, W. Kob,
\newblock  {\em J. Chem. Phys. } {\bf  140}, 164502 (2014)

\bibitem{w2}   W. Kob, L. Berthier,
\newblock  {\em Phys. Rev. Lett.} {\bf  110}, 245702 (2013)

\bibitem{w3}   L. Berthier, W. Kob,
\newblock  {\em Phys. Rev. E} {\bf  85}, 011102 (2012)



 \bibitem{cage} V. Teboul, M. Saiddine, J.M. Nunzi, J.B. Accary, 
 \newblock \emph{J. Chem. Phys. } \textbf{134}, 114517 (2011)

 
\bibitem{md3} S. Chaussedent, V. Teboul, A. Monteil,
 \newblock  {\em  Curr. Opin. Solid State Mater. Sci.} {\bf 7}, 111 (2003)

\bibitem{md4} V. Van Hoang, 
 \newblock  {\em  Philos. Mag.} {\bf 91}, 3443 (2011)
 
\bibitem{md5} D. Limmer, D. Chandler, 
 \newblock  {\em  J. Chem. Phys.} {\bf 135}, 134503 (2011)

\bibitem{md6} B. Kezic, A. Perera,
 \newblock  {\em  J. Chem. Phys.} {\bf 137}, 014501 (2012)


 
\bibitem{md7} V. Teboul, Y. Le Duff,
 \newblock  {\em  J. Chem. Phys.} {\bf 107}, 10415 (1997)
 
 
 \bibitem{potKA1} W. Kob, H.C. Andersen, 
 \newblock \emph{Phys. Rev. Lett.} \textbf{73}, 1376 (1994)

 \bibitem{potKA2} W. Kob, H.C. Andersen, 
 \newblock \emph{Phys. Rev. E} \textbf{51}, 4626 (1995)

 \bibitem{potG} M. Engel, H.R. Trebin, 
 \newblock \emph{Phys. Rev. Lett.} \textbf{98}, 225505 (2007)
 
 \bibitem{potDz1} M. Dzugutov, 
 \newblock \emph{Phys. Rev. A} \textbf{46}, R2984 (1992)
 
\bibitem{potDz2} J.P.K. Doye, D.J. Wales, F. Zetterling, M. Dzugutov, 
 \newblock \emph{J. Chem. Phys. } \textbf{118}, 2792 (2003)
 
\bibitem{potDz3}  M. Elenius, T. Oppelstrup, M. Dzugutov, 
 \newblock \emph{J. Chem. Phys. } \textbf{133}, 174502 (2010)

 
\bibitem{potKL1} A.J. Moreno, S.H. Chong, W. Kob, 
\newblock \emph{J. Chem. Phys.} \textbf{123}, 204505 (2005)

\bibitem{potKL2} S. Kammerer, W. Kob, R. Schilling, 
\newblock \emph{Phys. Rev. E} \textbf{58}, 2141 (1998)


\bibitem{pc1} H.J. Woo, X. Song,
\newblock \emph{J. Chem. Phys.} \textbf{116}, 4587 (2002)


\bibitem{pc2} R. Ni, M. Dijkstra,
\newblock \emph{J. Chem. Phys.} \textbf{134}, 034501 (2011)

\bibitem{pc3} M. Marechal, M. Dijkstra, 
\newblock \emph{Soft Mat.} \textbf{7}, 1397 (2011)

\bibitem{M3} M. Marechal, M. Dijkstra, 
\newblock \emph{Phys. Rev. E} \textbf{77}, 061405 (2008)

\bibitem{M1} C. Vega, P.A. Monson, 
\newblock \emph{J. Chem. Phys.} \textbf{107}, 2696 (1997)

\bibitem{M2} C. Vega, E.P.A. Paras, P.A. Monson,
\newblock \emph{J. Chem. Phys.} \textbf{96}, 9060 (1992)


\bibitem{M4} C. Vega, E.P.A. Paras, P.A. Monson,
\newblock \emph{J. Chem. Phys.} \textbf{97}, 8543 (1992)





\bibitem{berendsen} H.J.C. Berendsen, J.P.M. Postma, W. Van Gunsteren, A. DiNola, J.R. Haak
 \newblock  {\em  J. Chem. Phys.} {\bf 81}, 3684 (1984)



\bibitem{angell1} C.A. Angell,
 \newblock \emph{Science} \textbf{267}, 1924 (1995)
 
 
\bibitem{angell2} P.G. Debenedetti, F.H. Stillinger,
 \newblock \emph{Nature} \textbf{410}, 259 (2001)
 



\bibitem{stokes-einstein} Z. Shi, P.G. Debenedetti, F.H. Stillinger, \emph{J. Chem. Phys.} \textbf{138},
12A526 (2013)


\bibitem{cageM} M. Goldstein, 
 \newblock \emph{J. Chem. Phys.} \textbf{51}, 3728 (1969)









\end{thebibliography}
\end{document}